\documentclass[conference,10pt]{IEEEtran}

\usepackage[T1]{fontenc}
\usepackage{url}
\usepackage{cite,xcolor}
\usepackage[cmex10]{amsmath} 

\usepackage[cmintegrals]{newtxmath}
\usepackage{bm} 
\usepackage{graphicx}
\usepackage{enumitem}
\usepackage[normalem]{ulem}
\usepackage{microtype}

\graphicspath{{.}{../pictures/}{pictures/}{../../pictures/}}

\definecolor{awesome}{rgb}{1.0, 0.13, 0.32}

\newtheorem{theorem}{Theorem}
\newtheorem{lemma}{Lemma}
\newtheorem{example}{Example}
\newtheorem{remark}{Remark}
\newtheorem{corollary}{Corollary}

\usepackage{etoolbox}
\makeatletter
\patchcmd{\@begintheorem}{\textit}{\textbf}{}{}
\patchcmd{\@begintheorem}{\itshape}{\bfseries}{}{}
\makeatother

\begin{document}
\title{Very Pliable Index Coding} 


\author{%
  \IEEEauthorblockN{Lawrence Ong}
  \IEEEauthorblockA{The University of Newcastle\\
    Email: lawrence.ong@newcastle.edu.au}
  \and
  \IEEEauthorblockN{Badri N.\ Vellambi}
  \IEEEauthorblockA{University of Cincinnati\\
    Email: badri.vellambi@uc.edu}
}


\maketitle

\begin{abstract}
In the pliable variant of index coding, receivers are allowed to decode any new message not known a priori. Optimal code design for this variant involves identifying each receiver's choice of a new message that minimises the overall transmission rate. This paper proposes a formulation that further relaxes the decoding requirements of pliable index coding by allowing receivers to decode different new messages depending on message realisations. Such relaxation is shown to offer no rate benefit when linear codes are used, but can achieve strictly better rates in general. Scenarios are demonstrated for which the transmission rates are better when the message size is finite than when it is asymptotically large. This is in stark contrast to traditional communication setups.
\end{abstract}

\section{Introduction}

Traditional communication systems involve sending specific messages to specific receivers. Such requirements have been relaxed in modern applications like recommender systems, sensor networks, and dataset transmission for distributed learning. In these applications, the receiver's requirement is to obtain any \emph{new} message that it does not already have; there is no constraint on which new message is to be decoded.

Yet, codes designed for traditional communications are still being used for such applications. This paper explores the fundamental limits of communications with decoding pliability under the framework of index coding. Index coding~\cite{baryossefbirk11,arbabjolfaeikim18trends,ong2017} consists of one transmitter sending specific messages to multiple receivers through a common broadcast, where each of the receivers already has some subset of messages. Although the setup of index coding seems overly simplified, it has been shown to be equivalent (in terms of transmission rates and code design) to network coding, which is a general multi-source, multi-sink, multi-link network problem~\cite{effrosrouayheblangberg15,ongvellambiklieweryeoh21}.

Decoding pliability under the framework of index coding has been coined \emph{pliable index coding}~\cite{brahmafragouli15}. It has been shown that a code designed for pliable index coding achieves significant transmission rate improvement compared to its non-pliable counterpart (index coding).
For a problem with $n$ receivers each having a randomly selected subset of messages, such relaxation has been shown to decrease the required size of the broadcast codeword from $\sqrt{n}$ to $\log n$~\cite{brahmafragouli15}.

Pliable index coding shares much similarity with index coding, especially in code design. Index coding is equivalent to pliable index coding with a fixed decoding choice for each receiver. Thus, evaluating index coding with all possible decoding choices yields an optimal solution for pliable index coding, including the optimal code. While it is NP-hard to determine the best decoding choice, they have been found for a few special cases~\cite{liutuninetti20,ongvellambikliewer2019,OngConf3}. Lower bounds on the rates are derived using decoding-chain arguments, and achievability, using MDS codes or uncoded transmissions.

Although pliable index coding allows each receiver to decode any new message, the decoding choice must be fixed in the code design for all message realisations. This fixed-choice constraint is often unnecessary in the true spirit of decoding pliability.
In this paper, we study a relaxed version of pliable communications where the receivers are not tied to a decoding choice, but are free to decode different new messages upon receiving different transmissions. We call such a setup \emph{very-pliable} (VP) index coding.

Such a deviation from traditional decoding models complicates analyses, as many entropy-based information-theoretic tools no longer apply. For example, decoding rules of the form $H(X_i|Y,X_\mathcal{S}) = 0$ do not hold; here $i$ is the choice of the index of the message to be decoded, $Y$ is the received transmission, and $X_\mathcal{S}$ are the messages that the receiver has. Also, due to the absence of a fixed decoding choice, the concatenation of two (or more) very-pliable (VP) index codes does not yield a longer VP code.

\subsection{Contributions}

This paper establishes the following results:
\begin{enumerate}
    \item The broadcast rate for VP index coding (or VP rate, in short) is at least one. The broadcast rate is the main performance metric, given by the transmitted codeword length normalised to the message length.
    \item When restricted to linear encoders, the optimal VP rate is the same as optimal pliable index coding rate.
    \item There exist scenarios with finite message alphabets where
    \begin{enumerate}
        \item The VP rate is strictly lower (better) than that for the corresponding pliable setting.
        \item The VP rate is strictly lower than the asymptotic VP rate (as the alphabet size grows unbounded).
    \end{enumerate}
    \item A procedure to construct VP codes for a larger alphabet size from VP codes for smaller alphabets.
\end{enumerate}

Results~1 and 3 are obtained by restricting the number of messages that can be encoded to one particular codeword due to the decoding requirements, and then using a counting argument to lower bound the number of codewords in total. Achievability is obtained by constructing a hypergraph representation of the problem and evaluating the covering number of the hypergraph. Result~2 is obtained by using the solvability of a single  variable in a system of linear equations. Result~4 is obtained by concatenating a VP code with an MDS code.

Although it has been shown that linear codes can be strictly sub-optimal for some index-coding instances, examples of such instances and codes have not been reported. In this paper, we present very-pliable index-coding settings and their rate-optimal non-linear VP codes, which strictly outperform linear codes.

\section{Problem Formulation}\label{sec:formulation}
We use the following notation: 
$\mathbb{Z}^+$ denotes the set of natural numbers, 
$[a:b] := \{a, a+1, \dotsc, b\}$ for $a,b\in\mathbb{Z}^+$ such that $a < b$, and 
$X_S = (X_i: i \in S)$ for some ordered set $S$.

Consider a sender with $m \in \mathbb{Z}^+$ messages denoted by $X_{[1 : m]} = (X_1, \dots, X_m)$. Each message $X_i \in [0:k-1]$, where $k \geq 2$ denotes the message alphabet size. There are $n$ receivers having distinct subsets of messages, which we refer to as side information. Each receiver is labelled by its side information, i.e., the  receiver that has messages $X_{H}$, for some $H \subsetneq [1 : m]$, will be referred to as receiver $H$.

A problem instance is characterised by $(m,\mathbb U)$, where $m$ is the number of messages, and the set $\mathbb{U} \subseteq 2^{[1:m]} \setminus \{[1:m]\}$ represents the receivers in the instance. Given a problem instance $(m,\mathbb{U})$, a very-pliable index code (or VP code in short) of size $t \in \mathbb{Z}^+$ for message size $k\in\mathbb{Z}^+$ consists of
\begin{itemize}
\item sender's encoding function $\mathsf{E}: [0:k-1]^m \rightarrow [0:
t-1]$;
\item for each receiver $H\in\mathbb{U}$, a pair of decoding functions 
\begin{align*}
    &\mathsf{I}_H: [0:t-1] \times [0:k-1]^{|H|} \,\,\,\rightarrow [1:m]\setminus H\\
    &\mathsf{G}_H: [0:t-1] \times [0:k-1]^{|H|} \rightarrow [0:k-1]
\end{align*}
such that for any realisation $x_{[1:m]}\in [1:k-1]^m$, receiver~$H$ decodes a new message with index $\mathsf{I}_H(\mathsf{E}(x_{[1:m]},x_H)$ without any error. By new, we mean that for any $H\in\mathbb{U}$ and $x_{[1:m]}$, $\mathsf{I}_H(\mathsf{E}(x_{[1:m]}),x_H)\notin H$. The value of the message decoded by receiver~$H$ is given by $\mathsf{G}_H(\mathsf{E}(x_{[1:m]}),x_H)$. 
\end{itemize}

In pliable index coding, the index of the decoded message depends only on the receiver, and not the message realisation; in VP index coding, the message index \emph{can} vary with message realisation. Hence, the need for an additional decoding function $\mathsf{I}_H$ that specifies which message is decoded. The following example illustrates this key idea of \emph{very pliability}.

\begin{figure}[t]
\centering
    \includegraphics[width=8cm]{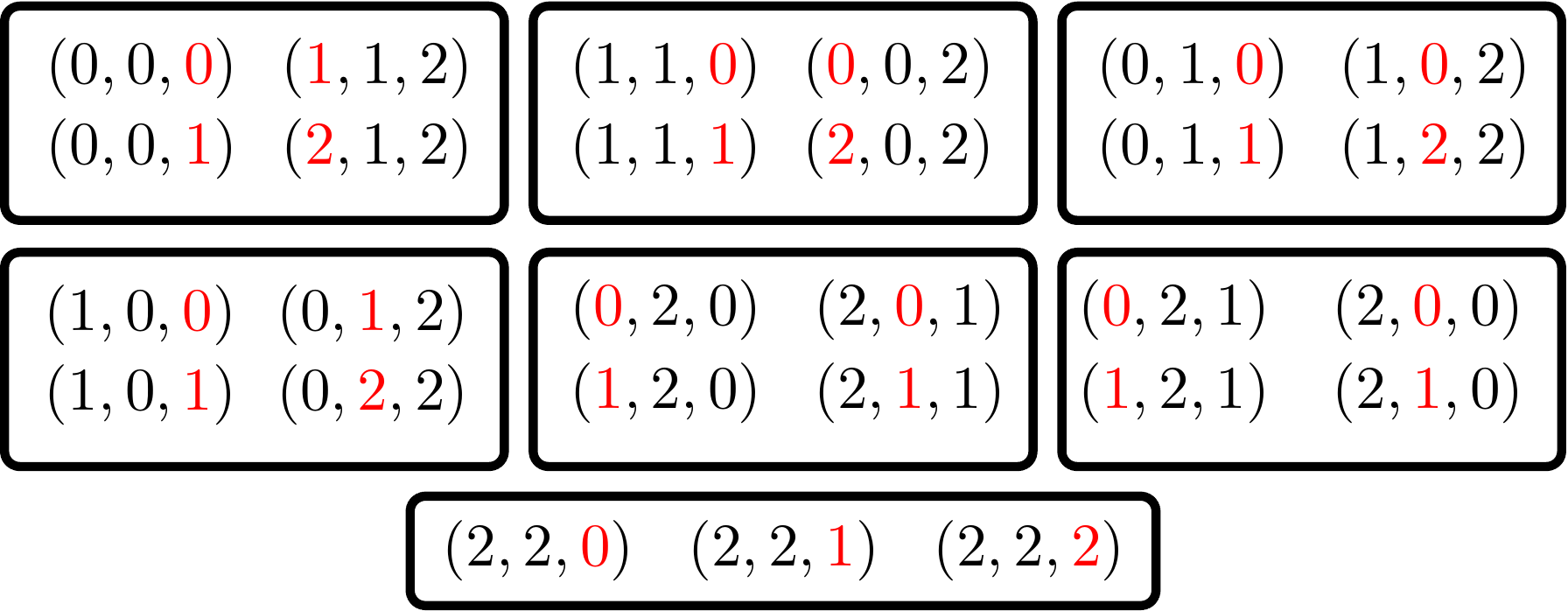}
\caption{A rate-optimal VP for $m=3$, $\mathbb{U} = \big\{ \{1\}, \{2\}, \{3\} \big\}$, and $k=3$.}
\label{fig:eg:0}
\end{figure}

\begin{example}\label{eg:0}
Consider the problem instance $m=3$, $\mathbb{U} = \big\{ \{1\}, \{2\}, \{3\} \big\}$. For $k=3$, a VP code for this instance is given in Figure~\ref{fig:eg:0}, where each box indicates all message realisations mapped to a particular codeword. Suppose that the codeword given by top left box is conveyed by the encoder. Then, receiver with side information $X_2$ will decode $X_1=0$ when $X_2=0$, and when $X_2=1$, the same receiver will decode $X_3=2$. Thus, in VP coding, a receiver can decode different messages for different received codeword and side-information realisations.
\end{example}

The aim is to find the optimal rate for a particular message size $k$, denoted by
\begin{equation}
    \alpha_k := \min_{\mathsf{E}, \{\mathsf{I}_H,\mathsf{G}_H\}} \frac{\log t}{\log k},
\end{equation}
the optimal rate over all $k$, denoted by
\begin{equation}
    \alpha_* := \inf_k \alpha_k,
\end{equation}
and the asymptotic rate, denoted by
\begin{equation}
    \alpha_\infty := \liminf_{k \rightarrow \infty} \alpha_k.
\end{equation}

For any problem instance $(m,\mathbb{U})$, a pliable index code differs from a VP code only in $\mathsf{I}_H$, where it is a constant function. So, a pliable index code is also a VP code, and
\begin{align}
    \alpha_k &\leq \beta_k, \quad \forall k \geq 2,\label{eq:rate-1-compare}\\
    \alpha_* &\leq \beta_{*},\\
    \alpha_\infty &\leq \beta_{\infty},\label{eq:rate-3-compare}
\end{align}
where $\beta$ denotes the corresponding rates for pliable index codes for the same problem instance.

For index codes and pliable index codes, we always have $\beta_{*} = \beta_{\infty}$~\cite{ong2017}. However, for some problem instances, we will show that not only can $\alpha_*$ and $\alpha_\infty$ be distinct, but
\begin{equation}
     \alpha_* < \alpha_\infty.
\end{equation}

Without loss of generality, the side-information sets $H$ of the receivers are distinct; all receivers having the same side information can be satisfied if and only if any one of them can be satisfied. Also, no receiver has side information~$H = [1:m]$ because this receiver cannot be satisfied. 

\section{Results}

\begin{theorem}\label{thm:alpha-grater-than-1}
For any problem instance and for any $k$, $\alpha_k \geq 1$.
\end{theorem}
This can be proven easily (using entropy) for index codes and pliable index codes, but not so for VP codes.

\begin{IEEEproof}[Proof of Theorem~\ref{thm:alpha-grater-than-1}]
Our proof bounds the number of message realisations that can be encoded to a codeword. Consider the encoding function $\mathsf{E}$ of any VP code. Let the set of message realisations that are encoded to a codeword $c$ by $\mathsf{E}^{-1}(c) \subseteq [0:k-1]^m$.

Consider an arbitrary receiver~$H$. Let the subset of $\mathsf{E}^{-1}(c)$ with a specific $x'_H$ as the receiver's side information be
\begin{equation}
    \mathcal{S}_H(c,x'_H) := \{ x_{[1:m]} \in \mathsf{E}^{-1}(c) : x_H = x'_H\}. \label{eq:si-partition}
\end{equation}
The decoding requirement of receiver~$H$ dictates that for any $x'_H$ and $c$, there is an index $i \in [1:m] \setminus H$ and value $v\in[0,k-1]$ such that for any $x_{[1:m]}\in \mathcal{S}_H(c,x'_H)$, $x_i=v$, i.e., the component of the $i^{\textrm{th}}$ message in every realisation in $\mathcal{S}_H(c,x'_H)$ is the same. Note that $i$ and $v$ are allowed to vary with $c$ and $x'_H$. This means there can be at most $k^{m-|H|-1}$ distinct message realisations in $\mathcal{S}_H(c,x'_H)$. Then, since $\mathsf{E}^{-1}(c)$ is the disjoint union of $\mathcal{S}_H(c,x'_H)$ for various $x'_H$,  we see that 
\begin{align} \label{eq:codeword-size}
\big|\mathsf{E}^{-1}(c)\big| = \bigcup_{x'_H} \big|\mathcal{S}_H(c,x'_H)\big| \leq k^{|H|}k^{m-|H|-1} \leq k^{m-1}.
\end{align}
Thus, at most $k^{m-1}$ message realisations can be encoded to each codeword. Since each message realisation $x_{[1:m]} \in [0:k-1]^m$ must be encoded to a codeword, the total number of codeword required $t \geq k^m/k^{m-1} = k$. Hence,  $\alpha_k \geq 1$.
\end{IEEEproof}

\begin{theorem}\label{thm:linear}
Consider any linear VP index code, that is, 
\begin{equation}
    \mathsf{E}(x_{[1:m]}) = \boldsymbol{E} \boldsymbol{x},
\end{equation}
where $\boldsymbol{E}$ is an $T \times m$ encoding matrix over $\mathbb F_q$, the finite field of size $q$, and $\boldsymbol{x}$ is an $m\times 1$ vector over $\mathbb{F}_q$ denoting the message realisation. Then
\begin{align}
        \alpha_q^\text{linear} &= \beta_q^\text{linear}, \quad \text{for any $q$},\label{eq:thm2-1}\\
    \alpha_*^\text{linear} &= \beta_*^\text{linear}, \label{eq:thm2-2}\\
    \alpha_\infty^\text{linear} &= \beta_\infty^\text{linear}.\label{eq:thm2-3}
\end{align}
The superscript {\upshape linear} denotes optimal rates over linear codes.
\end{theorem}

\begin{IEEEproof}[Proof of Theorem~\ref{thm:linear}]
Let $\boldsymbol{E}$ be the encoding matrix of a linear VP index code. Let $H$ be a receiver in the problem. The encoding operation can be written as
\begin{equation}
   \boldsymbol{c}= \boldsymbol{E}\boldsymbol{x} = \boldsymbol{E}_{H^c} \boldsymbol{x}_{H^c} + \boldsymbol{E}_H \boldsymbol{x}_{H},
\end{equation}
where $\boldsymbol{E}_{H^c}$ is the $T \times |H^c|$ submatrix corresponding to columns of $H^c$, and $\boldsymbol{x}_{H^c}$ is a $|H^c|\times 1$ vector corresponding to messages whose indices lie in $H^c$; similarly, $\boldsymbol{E}_{H}$ is the $T \times |H|$ submatrix corresponding to columns of $H$, and $\boldsymbol{x}_{H}$ is a $|H|\times 1$ vector corresponding to messages whose indices lie in $H$.

Let us suppose that the receiver $H$ receives codeword $\boldsymbol{c}$ from the encoder, and side information $\boldsymbol{x}_H$. As before, let us define
\begin{equation}
    \mathcal{S}_H(\boldsymbol{c}, \boldsymbol{x}_H) = \left\{ (\tilde{\boldsymbol{x}}_{H^c},\tilde{\boldsymbol{x}}_{H}): \begin{array}{c} \boldsymbol{E}_{H^c} \tilde{\boldsymbol{x}}_{H^c} + \boldsymbol{E}_H \tilde{\boldsymbol{x}}_{H}= \boldsymbol{c} \\ \tilde{\boldsymbol{x}}_{H}=\boldsymbol{x}_{H}     \end{array}\right\}.
\end{equation}
Note that a message realisation is in $\mathcal{S}_H(\boldsymbol{c}, \boldsymbol{x}_H)$ if and only if it is a solution to 
\begin{equation}
\left[\begin{array}{c|c} \boldsymbol{E}_{H^c} & \boldsymbol{E}_H \\ \hline \mathbf{0}_{|H|\times |H^c|} & \mathbf{I}_{|H|\times |H|} \end{array}  \right]\begin{bmatrix} \boldsymbol{X}_{H^c} \\ \boldsymbol{X}_{H} \end{bmatrix} = \begin{bmatrix} \boldsymbol{c} \\ \boldsymbol{x}_{H} \end{bmatrix}.
\end{equation}
Therefore, it follows that a message realisation is in $\mathcal{S}_H(\boldsymbol{c}, \boldsymbol{x}_H)$ if and only if it is a solution to 
\begin{equation}
\left[\begin{array}{c|c} \boldsymbol{E}_{H^c} & \mathbf{0}_{T\times |H|} \\ \hline \mathbf{0}_{|H|\times |H^c|} & \mathbf{I}_{|H|\times |H|} \end{array}  \right]\begin{bmatrix} \boldsymbol{X}_{H^c} \\ \boldsymbol{X}_{H} \end{bmatrix} = \begin{bmatrix} \boldsymbol{c}-\boldsymbol{E}_H\boldsymbol{x}_{H} \\ \boldsymbol{x}_{H} \end{bmatrix}. \label{eqn-affinesoln}
\end{equation}
From elementary matrix analysis we know the following:

\textit{In a consistent system of equations $\boldsymbol{Ax}=\boldsymbol{b}$, a variable $x_j$  has a unique solution if and  only if a linear combination of the rows of $\boldsymbol{A}$ yields $\boldsymbol{e}_j$, the binary vector with a single one in the $j$-th component.}

Therefore, we conclude from \eqref{eqn-affinesoln} that the receiver $H$ will decode a message successfully after receiving codeword $\boldsymbol{c}$ from the encoder, and using side information $\boldsymbol{x}_H$ if and only if it can decode the same message always, i.e., for every $x_{[1:m]}$.

Therefore, there exists a message index $j\in H^c$ that depends only on $\boldsymbol{E}$ whose (message) value will always (i.e., for any $x_{[1:m]}$) be identified correctly by receiver $H$. Since the problem and the receiver $H$ are arbitrary, it follows that every VP linear code has an equivalent pliable code with the same encoder, which then naturally yields the claim of the theorem.
\end{IEEEproof}
%
%
%
%

\begin{theorem}\label{thm:3}
There exist problem instances where 
\begin{equation}
    \alpha_k < \beta_k,
\end{equation}
for some $k\in\mathbb Z^+$.
\end{theorem}

\begin{IEEEproof}[Proof of Theorem~\ref{thm:3}]
Consider the problem instance in Example~\ref{eg:0} and the corresponding VP code presented in Figure~\ref{fig:eg:0}. Note that the rate of this VP code is $1.7712$, whereas for the pliable codes for this instance, $\beta_k = 2$ for all $k$~\cite{liutuninetti20}. 
\end{IEEEproof}

The code illustrated in Figure~\ref{fig:eg:0} was, in fact, shown to be optimal by exhaustive search on a certain coding hypergraph created as follows:
\begin{itemize}
    \item The vertex set $\mathcal{V} = [0:k-1]^m$ consists of all $k^m$ message realisations.
    \item The hypergraph contains a hyperedge $\mathsf{e}\subseteq \mathcal{V}$ if and only if $\mathsf{e}$ is a maximal subset of message realisations (i.e., vertices of the hypergraph) that can mapped to a codeword by the encoder while ensuring successful recovery of a new message by each receiver. For the problem instance in Example~\ref{eg:0}, $\{(0,0,0), (0,0,1), (1,1,2), (2,1,2)\}$ is a hyperedge, as adding any realisation to this set will violate the decoding requirement of one or more of the receivers. 
\end{itemize}

Any hyperedge in this coding hypergraph can be used to construct a codeword for a VP code. Therefore, the problem of designing a VP code for a problem reduces to identifying a collection of hyperedges in the hypergraph covering all vertices (i.e., message realisations). It then follows that the design of rate-optimal VP code reduces to the identification of minimal vertex cover of the underlying coding hypergraph. 

The following technicality arises in formulating the rate-optimal VP code from a minimal vertex cover: \textit{multiple hyperedges may cover a vertex}; however, a realisation can be assigned to only one codeword. To address this, we simply enumerate the hyperedges in any order, and assign message realisations to the first hyperedge they appear in. 

\begin{figure}[t]
\centering
    \includegraphics[width=8cm]{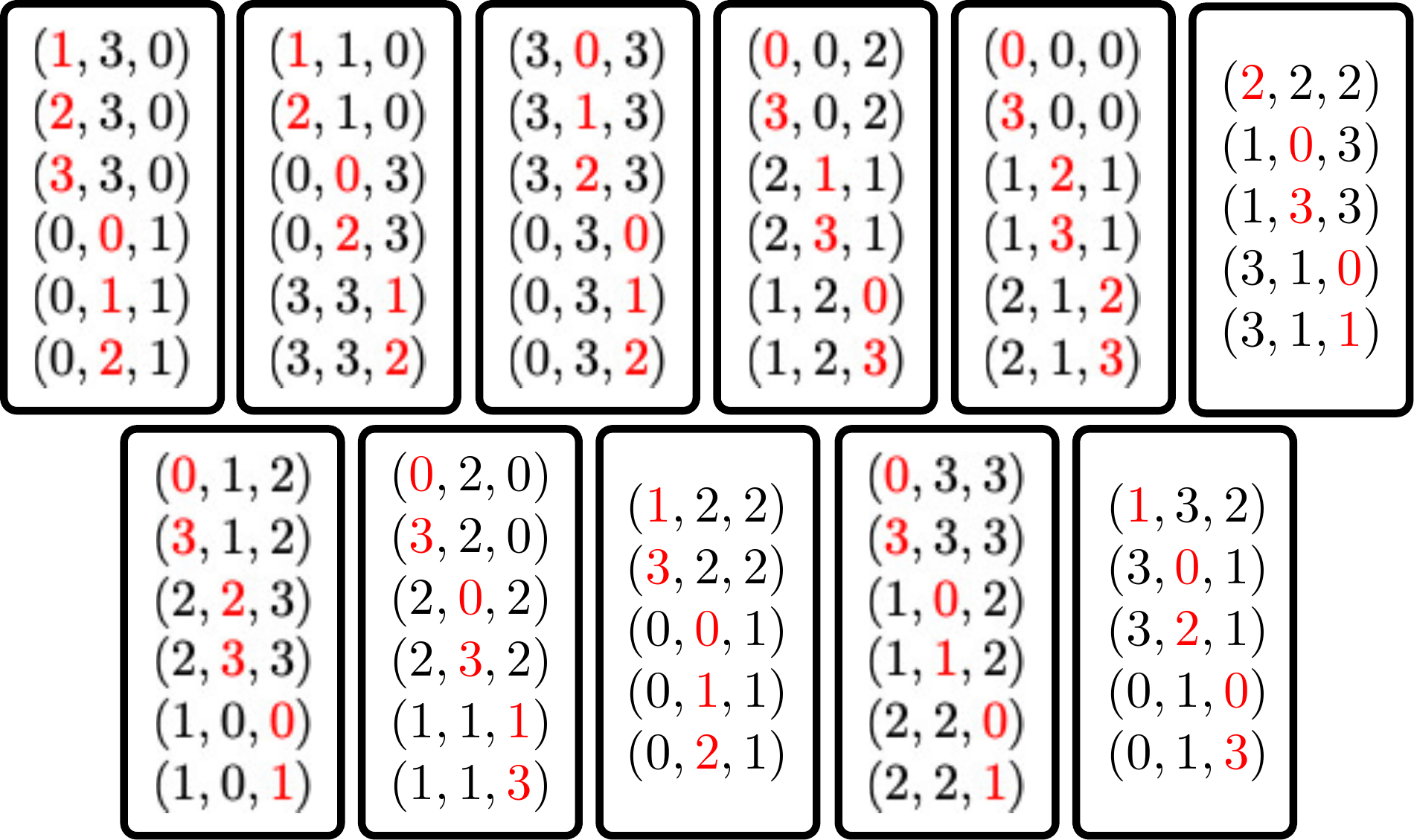}\\
\caption{A rate-optimal VP code for $m=3$,  $\mathbb{U} = \big\{ \{1\}, \{2\}, \{3\} \big\}$, and $k=4$.}
\label{fig:minimal-covering}
\end{figure}

From the above description, it is clear that $\alpha_k$ equals the \textit{covering number}~\cite[p.~1]{ScheinermanUllman} of the coding hypergraph. Through an implementation of the above hypergraph approach, we identified optimal rates for the problem instance $m=3, \mathbb{U} = \big\{ \{1\}, \{2\}, \{3\} \big\}$ for the following message alphabet sizes:
\begin{itemize}
    \item $k=2$: $\alpha_2 = 2 = \beta_2$
    \item $k=3$: $t=7$, and hence $\alpha_3 = 1.7712 < \beta_3$
    \item $k=4$: $t=11$, and hence $\alpha_4 = 1.7297 < \beta_4$
\end{itemize}
An optimal codebook for the last setup is shown in Figure~\ref{fig:minimal-covering}.

\begin{theorem}\label{thm:infty-equal}
Consider the problem instance with $m \geq 3$ messages and $\mathbb{U} = \big\{\{1\},\{2\},\ldots,\{m\}\big\}$. For this instance,
$\alpha_\infty=\beta_\infty=2$
\end{theorem}
\begin{IEEEproof}[Proof of Theorem~\ref{thm:infty-equal}]
This side-information setting belongs to the class of consecutive complete-$S$ problems, and for this setting, $\beta_\infty=2$~\cite{liutuninetti20}.
We only need to focus on $\alpha_\infty$. We will establish this result by bounding from above the maximum number of message realisations that can be mapped to any one coded message. To begin with, assume $k$ is the size of each message and let $\mathsf{E}: [0:k-1]^m\rightarrow [0:t-1]$ be a VP encoder for this problem. Let $c\in[0:t-1]$ and let us investigate the realisations in the pre-image $\mathsf{E}^{-1}(c)$. To assist our analysis, we classify the realisations in $\mathsf{E}^{-1}(c)$ as follows (see Figure~\ref{fig:structclass}):
\begin{itemize}
    \item Partition $\mathsf{E}^{-1}(c)$ into $m-1$ parts. Part $i\in[2:m]$ consists of all realisations in $\mathsf{E}^{-1}(c)$ where receiver~1\footnote{Recall that receiver~$H$ has $X_H$ as side information. When $H$ is a singleton, we call receiver~\{i\} as receiver~$i$.} decodes message $X_i$. Note that there is no Part 1.
    \item View the realisations in Part $i>1$ of the partition as a table with each row corresponding to the realisation and the columns ordered as $X_1$ as the first, $X_i$ as the second, and the rest in increasing order of message index starting from $X_{\min([2:m]\setminus \{i\})}$. For example in Part 2, the columns correspond to realisations of $X_1,X_2,X_3,\ldots, X_m$ in that order, whereas in Part 3, the columns correspond to realisations of $X_1,X_3,X_2,X_4,\ldots, X_m$ in that order. By construction, any realisation cannot appear in more than one partition.
    \item In Part $i>1$, we classify the rows (equivalently, realisations) into three cases as follows: 
    \begin{itemize}
    \item Case A: a row/realisation is of case A if for this realisation,
    receiver~1 decodes message $X_i$, and receiver~$i$ decodes message $X_\ell$ for some $\ell\notin\{1,i\}$.
        \item Case B: a row/realisation is of case B if for this realisation,
    receiver~1 decodes $X_i$, and receiver~$i$ decodes $X_1$, and receiver~$\min([2:m]\setminus \{i\})$ (i.e., the receiver that has the message whose index corresponds to the third column) decodes a message other than $X_1$ or $X_i$.
    \item Case C: a row/realisation is of case C if for this realisation,
    receiver~1 decodes message $X_i$, and receiver~$i$ decodes $X_1$, and receiver~$\min([2:m]\setminus \{i\})$ decodes either $X_1$ or $X_i$.
\end{itemize}
\end{itemize}
 \begin{figure}[t]
\centering
    \includegraphics[width=8.5cm]{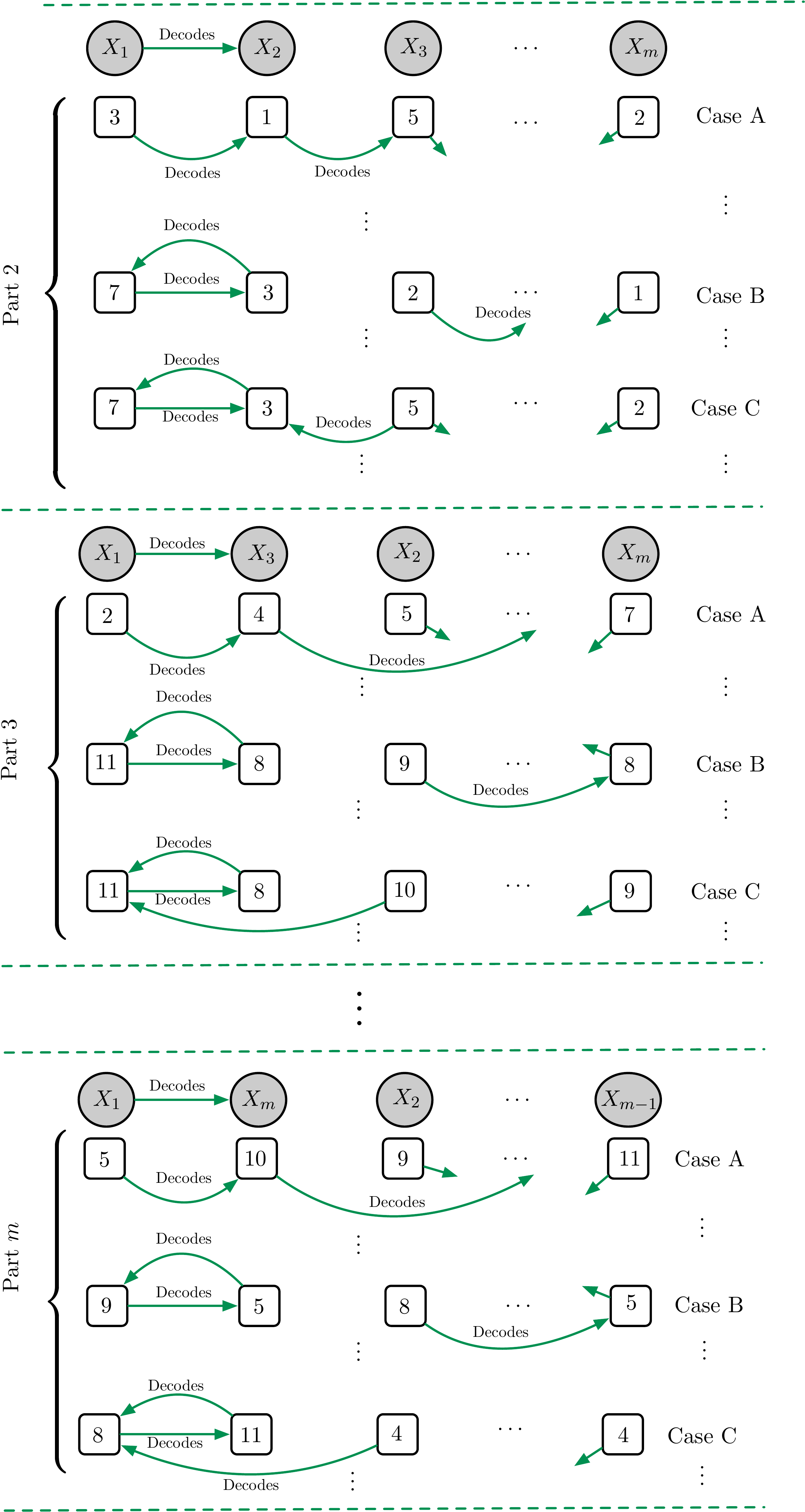}\\
\caption{Details of the partition of realisations mapped to a codeword $c$.}
\label{fig:structclass}
\end{figure}   
With this classification, we now bound the number of case A, B and C realisations in each pre-image $\mathsf{E}^{-1}(c)$.

Let us fix $a\in[0:k-1]$ and count the number of case-A realisations with $X_1=a$ in $\mathsf{E}^{-1}(c)$. The receiver~1 must decode some message $X_{j_a}$ whose value is, say $x_{j_a}$. By definition receiver~$j_a$ must decode another index $i_a\notin\{1,j_a\}$. Hence, the fact that a realisation is of case A, and $X_1=a$ determines uniquely two other messages indices and their values. Hence, the number of realisations of case A with $X_1=a$ is at most $k^{m-3}$ (that is, at most $k$ values for each of the remaining $m-3$ messages). Since $a\in[0:k-1]$, it follows that there can be at most $k\cdot k^{m-3} = k^{m-2}$ case-A realisations in  $\mathsf{E}^{-1}(c)$.

Let us fix $a,b\in[0:k-1]$. Let us count the number of case-B realisations with $X_1=a$ in $\mathsf{E}^{-1}(c)$. Receiver~1 must decode some message $X_{j_a}$ whose value is, say, $x_{j_a}$. By definition, receiver~$j_a$ must decode $X_1$ and receiver~$\min([2:m]\setminus \{j_a\})$ must decode a message that is neither $X_1$ nor $X_{j_a}$. Hence, the number of case-B realisations in $\mathsf{E}^{-1}(c)$  with $X_1=a$ and $X_{\min([2:m] \setminus \{j_a\})}=b$ is at most $k^{m-4}$ (i.e., at most $k$ values for each of the remaining $m-4$ messages). Since  $a,b\in[0:k-1]$, it follows that there can be at most $k^2 \cdot k^{m-4}=k^{m-2}$ case-B realisations in $\mathsf{E}^{-1}(c)$.

So far, we have upper bounded the total number of realisations of Cases~A and B in $\mathsf{E}^{-1}(c)$. For Case~C, we focus on each part of the partition shown in Figure~\ref{fig:structclass} individually. Let us fix $i\in[2:m]$ and $a\in[0:k-1]$, and count the number of Case-C realisations in Part $i$. Receiver~1 must decode some message $X_{i}$. By definition, receiver~$i$ must decode $X_1$, and receiver~$\min([2:m] \setminus \{i\})$ must decode a message that is either $X_1$ or $X_{i}$. So the value of message~$X_{\min([2:m]\setminus \{i\})}$ determines the values of both $X_1$ and $X_i$. Hence, the values of $X_1$ and $X_i$ are uniquely determined in any case-C realisation in Part $i$ when $X_{\min([2:m] \setminus \{i\})}=a$. This is only true in Part~$i$, unlike the counting arguments used for Cases~A and B, which apply to all of $\mathsf{E}^{-1}(c)$. So, there are at most $k^{m-3}$ Case-C realisations in Part $i$ with $X_{\min([2:m] \setminus \{i\})}=a$. Since $a \in [0:k-1]$ and $i \in [2:m]$, it follows that there are at most $k(m-1)\cdot k^{m-3}$ = $(m-1)k^{m-2}$ realisations of Case~C in $\mathsf{E}^{-1}(c)$.

Adding all the three cases, we see that there can be no more than $(m+1)k^{m-2}$ realisations in $\mathsf{E}^{-1}(c)$. Therefore, it follows that for any VP code over a message alphabet of size $k$,
\begin{align*}
     t \geq \frac{k^{m}}{\max_{c} |\mathsf{E}^{-1}(c)| }\geq \frac{k^{m}}{(m+1)k^{m-2}} = \frac{k^2}{m+1}.
\end{align*}
Since the above holds for any VP code and message alphabet size $k$, by limiting $k\rightarrow \infty$, we get the required result. 
\end{IEEEproof}

Combining Theorems~\ref{thm:3} and \ref{thm:infty-equal} for the problem instance $m=3$,  $\mathbb{U} = \big\{ \{1\}, \{2\}, \{3\} \big\}$, we get the following:
\begin{corollary}
There exist problem instances where
\begin{align}
    \alpha_* &< \alpha_\infty,\\
    \alpha_* &< \beta_*.
\end{align}
\end{corollary}

Note that unlike index codes or pliable index codes, VP codes cannot be concatenated, This is because although VP codes allow a decoder to decode any message---which can be different for each message realisation---if a message contains multiple parts, all decoded parts must be from the same message. Concatenating two VP codes may violate this decoding requirement as parts from different messages could be decoded. However, the following is possible:
\begin{theorem}\label{thm:concatenation}
Consider a problem instance where $|H| \geq 1$ for all $H \in \mathbb{U}$. Then, for any $k$,
\begin{equation}
   \alpha_{2k}\leq  \alpha_k\frac{\log k}{\log 2k} + (m-1) \frac{\log 2}{\log 2k}.
\end{equation}
\end{theorem}


\begin{IEEEproof}[Proof of Theorem~\ref{thm:concatenation}]
Consider any VP code of rate $\alpha_k$ that encodes messages of size $k$. By definition, such a code exists. For this code, it follows that $E(x)\in [0:k^{\alpha_k}-1]$ for $x \in [0:k-1]^m$.

Consider also a binary MDS code $E':\{0,1\}^m\rightarrow \{0,1\}^{m-1}$ defined by $E'(\boldsymbol{y}) = (y_1 \oplus y_2, y_2 \oplus y_3, \dots, y_{m-1} \oplus y_m)$, where $\boldsymbol{y} \in [0:1]^m$, and $\oplus$ is the modulo-2 addition. Note that for this code, a reciver having any one message can decode all other messages. 


To devise a VP code for message size $2k$, we proceed as follows. Instead of viewing the message alphabet as $[0:2k-1]$, let us view the message alphabet as $[0:k-1]\times\{0,1\}$. The encoder for the alphabet of size $2k$ will process messages over $[0:k-1]^m$ using $E$ and additionally process messages over $\{0,1\}^m$ using $E'$. This concatenated code has a total of $k^{\alpha_k} \cdot 2^{m-1}$ codewords. 

Each receiver for the concatenated code will use the decoder for $E$ to recover a message of size $k$, and use the decoder for $E'$ to recover the additional bit corresponding to the same message index decoded using the decoder for $E$. Note that the rate of this concatenated code is
\begin{equation}
    \frac{\log\big( k^{\alpha_k} 2^{m-1} \big)}{\log (2k)}.
\end{equation}
Since the concatenated code is a VP code that encodes messages with an alphabet size of $2k$, the claim follows.
\end{IEEEproof}

\begin{remark}
The concatenation technique used in the proof of Theorem~\ref{thm:concatenation} can be extended to the case where all receivers knows $p \in [0:m-1]$ messages as side information. In such case, for any $k$,
\begin{equation}
  \alpha_{kf(m,p)}\leq  \frac{\alpha_k\log k+(m-p)\log f(m,p)}{\log k+ \log f(m,p)) } ,
\end{equation}
where $f(m,p)$ is the minimum size of the finite field required to construct an $(m,m-p)$-MDS code.
\end{remark}

In the following, we establish problem instances where $\alpha_k=\beta_k$.
\begin{lemma}\label{cor:alpha-equals-beta}
For any problem instance where a rate-1 pliable index code exists for message alphabet size $k$,
\begin{equation}
    \alpha_{k^\ell} = \beta_{k^\ell} = 1,\label{eq:all-1}
\end{equation}
for all $\ell \in \mathbb{Z}^+$. 
\end{lemma}
\begin{IEEEproof}[Proof of Lemma~\ref{cor:alpha-equals-beta}]
Concatenate\footnote{Unlike VP codes, index codes and pliable index codes can be concatenated.} $\ell$ copies of the pliable index code to obtain $\beta_{k^\ell} \leq 1$ for any $\ell$. The result then follows from  \eqref{eq:rate-1-compare} and Theorem~\ref{thm:alpha-grater-than-1}.
\end{IEEEproof}

We now show that equality between $\alpha_k$ and $\beta_k$ can also hold for non-trivial cases where $\alpha_k > 1$.

\begin{lemma}\label{thm:alpha-equals-beta}
There exist problem instances where $\alpha_k = \beta_k > 1$ for all $k$.
\end{lemma}

\begin{IEEEproof}[Proof of Lemma~\ref{thm:alpha-equals-beta}]
Consider the problem instance with $m=3$ and $\mathbb{U} = \big\{ \emptyset, \{1\}, \{1,2\}, \{1,3\} \big\}$
Consider any codeword $c$ of a VP code of size $t\in\mathbb{N}$ for this instance, and fix $a \in [0:k-1]$. Consider the event that $X_1=a$. Receiver~$1$ (i.e., the receiver that has $X_1$) will use the codeword and the fact that $X_1=a$ to uniquely identify message, say $X_{i_a}$ as $x_{i_a}$, where $i_a\in\{2,3\}$. Now, since receiver~$\{1,i_a\}$ must also decode a message, it follows that the receiver~1 can use the decoded message $X_{i_a}=x_{i_a}$ to decode the value of the remaining message $X_{\{1,2,3\}\setminus\{1,i_a\}}$.

Hence, it follows that any $c$ and $a\in[0:k-1]$, the pre-image $\mathsf{E}^{-1}(c)$ can have at most one realisation with $X_1=a$. As $a \in [0:k-1]$, $|\mathsf{E}^{-1}(c)|\leq k$. Since this is true for any VP code, we must have $t \geq k^3 / k = k^2$. From \eqref{eq:rate-1-compare}, we note that  $\beta_k\geq \alpha_k \geq 2$. However, the pliable index code $\mathsf{E}(x_{[1:3]}) = (x_2,x_3)$ has a rate of $2$. Hence, $\alpha_k=\beta_k = 2$.
\end{IEEEproof}






\clearpage
\bibliography{otherpublications,mypublications}

\end{document}